\documentstyle[preprint,tighten,aps]{revtex}
\begin{document}
\draft
\title{Real CP Violation}
\author{A. Masiero}
\address{SISSA -- ISAS, Trieste and INFN, Sez. Trieste, Italy}
\author{T. Yanagida}
\address{Department of Physics and RESCEU, Univ. of Tokyo, 
Tokyo 113-0033, Japan}
\maketitle

\begin{abstract}
We propose a new mechanism called  ``real CP violation" to
originate
spontaneous CP violation. Starting with a CP conserving theory with scalar
fields in the adjoint representation of a global or local non-abelian
symmetry, we show that even though the VEV's of such scalars are real they
give rise to a spontaneous violation of CP. We provide an illustrative
example of how this new mechanism of CP violation can give rise to
physically significant phases which produce a complex CKM mixing matrix.
This mechanism may prove useful in string models with moduli in the
adjoint representation as well as in tackling the strong CP problem.

\end{abstract}


Since its experimental discovery in 1964, many mechanisms to originate CP
violation in K physics have been proposed. They can be grouped into two
classes: explicit and spontaneous CP violation. In the former case the
Lagrangian describing electroweak interactions contains some terms which are
not CP invariant. For instance, some Yukawa couplings may be complex
giving rise to a complex CKM matrix after diagonalization of the fermion
mass matrices. On the contrary, in the spontaneous option one starts with
a CP invariant Lagrangian, but the vacuum of the theory is not CP
invariant \cite{lee}. Typically one has some scalar fields developing complex
vacuum expectation values (VEV's) with some phases remaining after
exploiting all the invariances of the theory. These physical phases appear
in the quark mass matrices giving rise once again to a complex CKM matrix.

The possibility that CP is broken spontaneously is quite attractive.
Still lacking the underlying theory explaining the origin of the Yukawa
couplings, in the explicit case we introduce CP violation ``by hand" in
these complex couplings. Moreover if one decides that CP is not a good
symmetry of the theory since the beginning, one may expect arbitrarily
large violations of CP also in the strong interactions due to the presence
of the $\theta$ term in the QCD Lagrangian \cite{rebbi}. On the other hand,
the spontaneous breaking of CP by  the vacuum of the theory is
more linked to the ``dynamics" of the theory itself and, if CP is a good
symmetry to start with, the $\theta$ term has to be vanishing in the
initial Lagrangian \cite{strongcp}. Obviously this fact does not imply by
itself that the
strong CP problem \cite{peccei} is solved since the subsequent spontaneous
violation of
CP with phases in the quark mass matrices in general gives rise to an
effective $\theta$ which is too large.

Here we come back to the idea that CP is broken only spontaneously. We
propose a new mechanism for this breaking which does not entail the
request of having complex VEV's of the scalar fields. For this reason we
call it ``real" CP violation and we show that it can be generally applied
in theories with non-abelian global or gauge symmetries. The
key-ingredient is to have a set of scalars sitting in the adjoint
representation of these symmetries. Then, even if these scalars have real
VEV's (which is generally the case for the real fields in the adjoint), CP
is broken by the vacuum of the theory. 

Apart from the interest in itself of this new mechanism   for spontaneous CP
violation, we think that there are potentially relevant applications. In
particular it may give rise to a source of CP violation in string
theories with moduli in the adjoint representation \cite{bachas} and it can be
relevant
for the solution of the long-standing problem of strong CP violation
\cite{peccei}. We
will elaborate more on this latter aspect in the second part ot this Letter.

First we introduce the mechanism of ``real" CP violation. The way one
defines CP transformations in the presence of a non-abelian symmetry
presents an important difference with respect to the usual way CP is
defined in the abelian case, say in QED. For simplicity, consider an SU(2)
fermionic current coupled to the triplet of vector bosons $W_i$, $i=1,2,3$.
The demand that this interaction lagrangian be invariant under CP entails
that $W_3$ and $W_1$ transform into themselves, while $W_2$ has to go into
$-W_2$ under a CP transformation. This is equivalent to say that, having
defined $W^+$ and $W^-$ in terms of $W_1$ and $W_2$ in the usual way, CP
interchanges $W^+$ and $W^-$. Consider now that we replace the $W$ vector
bosons with an SU(2) triplet of real scalar fields $\phi$. Once again the
presence
of $\tau_2$ in the SU(2) generators with $(\tau_2)^T=-(\tau_2)$, implies
that under a CP transformation the second component $\phi_2$ of the scalar
triplet has to be odd if the interaction  respects CP invariance. Hence, a
VEV of this scalar component, although it is obviously real, leads to a
spontaneous breaking of CP. The key-point is that in the non-abelian case
some of the generators are anti-symmetric and the corresponding scalar
components of the adjoint representation have to be odd under CP if we
want to find a consistent definition of CP to have the interaction
lagrangian invariant under it.   

We now come to the second task of this Letter, namely we show that the
abovementioned mechanism of "real" CP violation can produce physical
phases which show up at the level of the CKM quark mixing matrix. To this
goal, we provide an illustrative example based on a horizontal $SU(3)_H$
symmetry which may be global or gauged. We introduce three scalar octets,
that we generically denote with $\phi$  and a singlet $\phi_0$. As
for fermions, consider the 2
vector-like triplets $U_{(L,R)}$ and $D_{(L,R)}$ which are singlets under
the $SU(2)$ of the standard model (SM) and triplets of the colour $SU(3)$
symmetry.  They can get a direct large mass $M_U$ and $M_D$, respectively.
The enforcement of CP violation ensures that these masses are real. 
 Let us now make the
connection to the low-energy part of the model with the usual $u$ and $d$ SM
quarks. Also $u$ and $d$ are triplets under $SU(3)_H$. Hence we can write
the Yukawa couplings of the right-handed components of $u$ and $d$ with
the left-handed components of the corresponding $U$ and $D$ and the above
$\phi$ fields. Since we ask for CP conservation all these couplings are
real. Notice that $u_R$ and $d_R$ have the same quantum numbers of $U_R$
and $D_R$. Since we want to avoid that the previous Yukawa terms put into
communication also the right-handed components of $U$ and $D$ with their
left-handed counterparts, we impose a discrete symmetry under which $u_R$,
$d_R$ and all the $\phi$ fields are odd, while $U$ and $D$ are even.
Finally we introduce also the usual SM Higgs doublet $H$. We now have the
new couplings of H with $U_R$, $D_R$
and $u_L$, $d_L$. Then, the tree level exchange of D gives rise to the
effective interactions :

\begin{equation}
L_{eff} = \frac{\bar{d_R} (g_d \phi^a \lambda^a + g_d^{'}\phi_0)q_LHh_d}{M_D},
\label{eq:eff}
\end{equation}
where $g_d$, $g_d^{'}$ and $h_d$ denote the Yukawa couplings with $\phi$,
$\phi_0$ and
$H$, respectively, $\lambda^a$ are the Gell-Mann matrices of
$SU(3)_H$, $M_D$ is the direct mass of $D$ and, finally,  $q_L$ is the usual
doublet of the left-handed up- and down-quarks.  Analogous contributions
to the up quark sector arise with the different Yukawa couplings $g_u$,
$g_u^{'}$ and
$h_u$ . When the scalar fields get a VEV, the above $L_{eff}$ produce mass
matrices for the up- and down-quarks which are hermitian. The presence of
three $\phi$ octets assure that all components, in particular those
related to the antisymmetric Gell-Mann matrices, get a nonvanishing VEV.
Hence the quark mass matrices possess three phases. It is easy to see that
one combination of them can never be reabsorbed by redefining the quark
fields. Thus the CKM phase appears.

The fact that the quark mass matrices although complex are hermitean
suggests that the ``real" CP violation may prove useful in tackling the
strong CP problem. Actually for the $\theta$ problem we need the full
quark matrix involving both the ordinary and the heavy new quarks $U$
and $D$. For instance, if we consider the down sector, we have the
following renormalizable interactions and mass matrix:

\begin{equation}
(\bar{d_R} \bar{D_R})
\left( \begin{array}{cc}
0 & \phi + \phi_0 \\
H & M_D
\end{array} \right)
(d_L D_L)^T,
\label{eq:mass}
\end{equation}
where the integration of the heavy $D_{R,L}$ fields induces 
$L_{eff}$ in eq.~(\ref{eq:eff}). $H$,
$\phi$ and $\phi_0$ 
denote the mass terms coming from the VEV's of $H$, $\phi$ and $\phi_0$, 
respectively. Notice that the  VEV of $H$  can
always be made real by performing a $U(1)$ hypercharge rotation on $H$.
Given the hermiticity of the matrix block $\phi$, we conclude that the
determinant of the above mass matrix  is real.
 
The
$\theta$ term of the QCD lagrangian
vanishes because of the initial CP invariance of the theory, while the
contribution to the effective $\theta$ arising from the rotation of quark
fields to bring them to the physical basis vanishes at the tree level
since it is proportional to the argument of the determinant of the above quark
mass matrix. The point now is that the phenomenologically
required
smallness of $\theta$ requires the quark mass matrix hermiticity to be
spoiled by very tiny effects \cite{strongcp}. This computation would require
the detailed
formulation of a model which is beyond the scope of this Letter. Here we
limit ourselves to a few comments on the possible suppression of the
contributions giving rise to a nonvanishing effective $\theta$.

The dangerous corrections spoiling the hermiticity of the quark mass
matrices arise from loop contributions involving the presence of quartic
terms in the $\phi$ fields as well as from terms of the kind ${\phi}^2
H H^*$. Having the scale of $SU(3)_H$ breaking large compared to the
electrowek scale, the couplings of the latter terms have to be small not
to create a hierarchy problem for the $H$ mass. If we ask also for the
quartic couplings to be small, we get relatively low masses for the $\phi$
scalars, for instance in the TeV region, where they can become accessible
in next hadronic accelerators. Notice that the demand that the
coefficients of the $\phi$ quartic terms  be small may be naturally
accomplished if the scale of the new physics is large enough to allow for
a substantial running of such couplings. 

After eliminating the danger represented by the above corrections
proportional to the quartic scalar terms, we are left with only the Yukawa
couplings for the fermions. However, such couplings at the loop level can
induce only the renormalization of the fermion wave functions.  Calling
$Z$ and $Z'$ such wave function renormalizations of the right- and
left-handed fields in eq.~(\ref{eq:mass}), we obtain:

\begin{equation}
(\bar{d_R} \bar{D_R})Z
\left( \begin{array}{cc}
0 & \phi+\phi_0 \\
H & M_D
\end{array} \right)
Z'(d_L D_L)^T.
\label{eq:Z}
\end{equation}

$Z$ and $Z'$ have to be hermitean. Hence the determinant of the whole
matrix in eq.~(\ref{eq:Z}) remains real and, thus, there is no
contribution to a non-vanishing $\theta$ from these terms. 

Another possibility for
suppressing the
hermiticity-breaking corrections could be to supersymmetrize the proposed
scheme. Then the radiative corrections to the fermion mass matrices would
be suppressed by at least two powers of the ratio of the scale of
low-energy SUSY breaking to the large scale of the theory. Corrections
leading to fermionic wave function renormalization do not enjoy such a
kind of protection, however, following the above mentioned argument,  we
conclude that they do not give rise to a non-vanishing $\theta$. 
However, in SUSY theories the chiral supermultiplets are complex even if
they belong to the adjoint representation and, hence, they may have
complex VEV's in general spoiling the hermiticity of the mass matrices. We
need some dynamical reason for them to take only real VEV's.
   
In conclusion, we have proposed the new mechanism of ``real" CP violation
to account for spontaneous breaking of CP in models with scalar fields in
the adjoint representation of some global or local non-abelian symmetry.
The mechanism allows for spontaneous CP violation even though no complex
VEV occurs. The resulting CP violating phases leak to the fermionic mixing
sector giving rise to a welcome complex CKM matrix.  We pointed out that this
idea may find interesting
applications in those string theories with moduli in the adjoint
representation as well as in tackling the strong CP problem in the context
of the spontaneous CP proposals. The illustrative example that we offered
shows that it may be of interest to pursue in this direction to build a
complete model of real CP violation in the non-SUSY or SUSY contexts. 

We thank W. Buchmueller for interesting discussions and the DESY Theory
Group for its kind hospitality. The work of A.M. is partially
supported by
the EEC TMR Project "BSM", Contract Number ERBFMRX CT96 0090.

\end{document}